\documentclass[journal,twoside,web]{ieeecolor}
\usepackage{tmi}
\usepackage{cite}
\usepackage{amsmath,amssymb,amsfonts}
\usepackage{graphicx}
\usepackage{textcomp,bm}
\usepackage{ulem}

\def\BibTeX{{\rm B\kern-.05em{\sc i\kern-.025em b}\kern-.08em
    T\kern-.1667em\lower.7ex\hbox{E}\kern-.125emX}}
\begin{document}
\title{Flexible Bayesian Support Vector Machines for Brain Network-based Classification}
\author{Jin Ming, Suprateek Kundu, \IEEEmembership{Member, IEEE}
\thanks{This manuscript was first submitted on May 16th, 2022. ``This work was supported in part by the NIH awards R01 AG071174 and R01MH120299'' }
\thanks{ J. Ming is at Emory University 1518 Clifton Road, Atlanta, GA 30322 (e-mail: jin.ming@emory.edu). }
\thanks{S. Kundu is at The University of Texas MD Anderson Cancer Center, Houston, TX 77030 USA (e-mail: skundu2@mdanderson.org).}}

\maketitle

\begin{abstract}
\underline{Objective}: Brain networks have gained increasing recognition as potential biomarkers in mental health studies, but there are limited approaches that can leverage complex  brain networks for accurate classification. Our goal is to develop a novel Bayesian Support Vector Machine (SVM) approach that incorporates high-dimensional networks as covariates and is able to overcome limitations of existing penalized methods. \underline{Methods:} We develop a novel Dirichlet process mixture of double exponential priors on the coefficients in the Bayesian SVM model that is able to perform feature selection and uncertainty quantification, by pooling information across edges to determine differential sparsity levels in an unsupervised manner.  We develop different versions of the model that incorporates static and dynamic connectivity features, as well as an integrative analysis that jointly includes features from multiple scanning sessions. We perform classification of intelligence levels using resting state fMRI data from the Human Connectome Project (HCP), and a second Attention Deficiency Hyperactivity Disorder (ADHD) classification task. \underline{Results:} Our results clearly reveal the considerable greater classification accuracy under the proposed approach over state-of-the-art methods. The multi-session analysis results in the highest classification accuracy in the HCP data analysis. \underline{Conclusion:} We provide concrete evidence that the novel Bayesian SVMs provides an unsupervised and automated approach for network-based classification, that results in considerable improvements over penalized methods and parametric Bayesian approaches. \underline{Significance:} Our work is one of the first to conclusively demonstrate the advantages of a Bayesian SVM in network-based classification of mental health outcomes, and the importance of multi-session network analysis.
\end{abstract}


\section{Introduction}
\label{sec:introduction}
\IEEEPARstart{T}{here} is increasing evidence about the potential of resting state functional Magnetic Resonance Imaging (rs-fMRI) in terms of informing diagnosis of different disease conditions. 
There are several approaches for constructing networks for brain functional connectivity (FC) [1] that rely on computing correlations between a set of pre-defined brain regions or nodes. When the correlations are time-averaged and stationary over the length of the fMRI session, these measures correspond to static connectivity; whereas more recent literature has also included dynamic connectivity models that allow time-varying correlations [2]. Moreover, the connectivity is typically computed in terms of pairwise correlations or partial correlations, with the latter being increasingly favored in the statistical and machine learning literature [3]. 
In an undirected network with $V$ nodes that is of interest in this article, the edge set can be fully represented by the corresponding $V(V-1)/2$ off-diagonal elements in a symmetric correlation matrix.

There is some literature on Bayesian linear regression models for prediction based on brain networks [4]. Overcoming the limitations of linear models, more recent work have focused on Bayesian non-linear regression models involving Gaussian processes [5], and some deep learning approaches have also been proposed [6][7]. While deep learning methods are attractive in terms of providing an end-to-end pipeline, they often lack interpretability, and require a large training samples to adequately fit a massive number of embedded model parameters. Classification approaches based on network edge features have been proposed in literature [8] that have been shown to result in improvements over approaches using global network metrics [9] or those that test for network features to find significant differences between two populations, with a multiple testing correction and without the need to construct a classifier at all [10]. Support vector machines (SVM) are perhaps the most widely used classifiers that have been successfully used in several disease areas including schizophrenia, bipolar disorder, autism spectrum disorder, ADHD, Alzheimer's disease, and so on. See the review article by [11].



It is to be noted that the overwhelming majority of SVM-based classification approaches using FC as covariates rely on penalized or optimization methods to derive point estimates for model fitting. While penalized approaches are often useful in practical applications, there are several limitations. First, the results under penalized SVM methods are often sensitive to the choice of the penalty parameter that is often tuned via cross-validation. The problem is exacerbated in brain network applications, since the choice of the penalty parameter reflects an overall sparsity level for the model, but it may not be adaptive to different levels of shrinkage across different subsets of network edges, resulting in inadequate performance. For example, in brain network applications, the variability in the mental health outcomes are often driven by a small proportion of significant edges with varying importance. Existing linear penalized approaches rely on a global penalty parameter, and may not be readily adaptive to differential sparsity levels. For example, the lasso penalty is typically known to result in inflated models involving an overly large set of estimated non-zero features [12]. Second, penalized SVM approaches fail to report measures of uncertainty, which is of paramount importance in high-dimensional neuroimaging applications. This is due to the fact that the brain network is derived from resting state fMRI (rs-fMRI) data via suitable algorithms, and hence are subject to measurement error and mis-specifications. In these scenarios, a point estimate may not suffice, and additional measures of uncertainty via credible intervals may be highly desirable. Third, inferring significant effects under existing penalized SVM methods often require computationally expensive procedures such as permutation tests [1] that may be challenging to implement for high dimensional network applications. As a result, penalized methods may yield inaccurate feature selection results.

To address the above challenges in this article, we propose a novel non-parametric Bayesian support vector machine approach for classification, based on high-dimensional brain networks. The key motivation behind the non-parametric Bayesian approach is to adaptively cluster the subsets of edges according to similarity in edge importance under a Dirichlet process (DP) mixture prior, with distinct clusters reflecting varying degrees of importance of the edges in the SVM model. By pooling information across subsets of edges and allowing for different degrees of Bayesian shrinkage, the proposed approach is able to estimate the edge-specific importance in a robust and accurate manner, which results in considerable improvements over existing approaches. The number of clusters and the cluster memberships for the edge importance weights are determined in a data adaptive manner, under a fully non-parametric Bayesian set-up. The method is scalable to high-dimensional networks and performs well in terms of feature selection and classification performance in these high-dimensional settings. The proposed method overcomes the limitations of penalized or optimization based SVM methods described earlier. For example, it is naturally able to quantify uncertainty for the parameter estimates via credible intervals, and it provides an integrated framework for inference required for feature selection. In addition, it has considerable advantages over parametric Bayesian approaches in terms of ensuring model parsimony, greater computational efficiency, and superior numerical performance.

The main contributions of this article are as follows. First, the proposed approach is one of the first Bayesian SVM classification approaches based on static brain networks that is able to adaptively estimate edge importance weights and infer significant network edges, resulting in improvements over penalized methods. 
Second, we extend the proposed approach to incorporate dynamic networks as features, which is able to naturally infer significant dynamic edges. Our analysis provides a direct comparison between the classification performance based on static versus dynamic resting state networks, which is of independent interest. 
Third, we extend the proposed method to integrate static networks from multiple fMRI sessions pertaining to the same set of samples. There is very limited work in neuroimaging literature for such integrative analysis, although recent studies involving penalized methods have indicated the benefits of performing multi-modal [13] and multi-task  [14]  analysis. To our knowledge, we propose one of the first Bayesian classification approaches for systematically integrating multi-session rs-fMRI networks. Fourth, we evaluate the effectiveness of our proposed framework via extensive numerical studies on various datasets that include rs-fMRI data from: (i) the Human connectome project (HCP) with the goal to classify intelligence levels; and (ii) a Attention-Deficit/Hyperactivity Disorder (ADHD) dataset, where the goal is to classify ADHD individuals versus neurotypical controls (NC). To our knowledge, our paper is one of the first to establish the utility of sophisticated Bayesian SVMs based on high-dimensional brain networks as a powerful classification tool in neuroimaging studies.



\section{Related Literature}
There is great interest in understanding the neural underpinnings of individual differences in intelligence, because it is one of the most important predictors of long-term life success. Intelligence may be measured via cognitive measures that may include fluid intelligence, defined as the ability to use inductive and deductive reasoning (independent of previously acquired knowledge) to solve new problems, or crystallized intelligence that involves knowledge that comes from prior learning and past experiences, among others. Such intelligence measures are assumed to be tied to brain structure; however, it is a major challenge to relate structural and functional properties of brain to complex behavioural expression or function [15].
 Existing literature has used neuroimaging-derived features such as whole brain volume, regional gray and white matter volumes or regional cortical volume/thickness and diffusion indices, which may smooth over discriminative features at a finer resolution and are often inadequate predictors of intelligence[16][17]
. A more recently emerging line of work has started to investigate prediction strategies for intelligence based on fMRI and derived features such as functional connectivity that has shown greater promise [18][19][20].

While intelligence prediction views cognition on a continuous scale, it may not be immediately successful in delineating the neural underpinnings of individual differences in intelligence. For example, [21] showed that functional connectivity profiles act as a `fingerprint' that can accurately identify subjects. In this paper we address a related question that investigates whether functional connectivity is able to distinguish individuals with varying levels of intelligence. Hence our objective can be also viewed as a type of connectome fingerprinting, where the goal is to identify subsets of individuals within a certain intelligence spectrum. Note that the investigation of optimal classifiers of cognitive levels or other mental health diagnosis, based on brain functional connectomes, is an open problem that requires greater attention [22], which we attempt to address in this article. 

Although classification of cognitive/intelligence levels provide a strong motivation for us to develop the proposed approach, the methodology developed in this article can be generalized to other mental health problems as well. In order to test the generalizability of the proposed classifier, we evaluate the performance on an independent dataset involving the problem of classifying individuals with ADHD versus neurotypical controls. This second application illustrates the strong versatility of the proposed approach for classifying mental health conditions that go beyond intelligence finger-printing. Such an analysis is well-founded based on existing evidence that suggests the utility of resting-state fMRI in modeling ADHD phenotypes [23]. 



\section{Methods and Materials}
\subsection{Description of Datasets}
\subsubsection{ HCP  study and pre-processing details}
The Human Connectome Project (HCP) contains resting state as well as task fMRI scans for adults, along with a battery of cognitive measurements including fluid intelligence and crystallized intelligence. The HCP developed a novel multimodal parcellation that defines brain regions based on a combination of structural and functional features [24] having 360 nodes that was used for our analysis. 
In the HCP data, each subject had two sessions of resting-state fMRI scans on the first day, where each scan was 14:33 minutes with TR=0.72, resulting in 1200 time points. Further, these scanning sessions were repeated on another follow-up visit. We used the cortical surface data from the FIX pre-processed left-right phase-encoding scan for our analysis (denoted as LR1 and LR2 corresponding to the first and second visits). The pre-processing pipeline included slice timing correction, rigid body re- alignment, and normalization to the EPI version of the MNI template. The time courses were temporally detrended in order to remove gradual trends in the data, and spatially coherent noise was removed and motion correction was performed using six rigid body realignment parameters and their first derivatives. Subsequently, the data was spatially smoothed with a 6mm FWHM Gaussian kernel and bandpass filtered be- tween 0.01-0.1 Hz. Finally, spike correction was performed 
as an alternative to motion scrubbing.  For more details of pre-processing, see [25].  

\subsubsection{ ADHD study and pre-processing details}
Pre-processed rs-fMRI data used was obtained from The Connectomics in NeuroImaging Transfer Learning Challenge (CNI-TLC) involving children. The CNI-TLC data was amassed retrospectively across multiple studies conducted by the Center for Neurodevelopmental and Imaging and Research (CNIR) at the Kennedy Krieger Institute (KKI) in Baltimore, Maryland. The overall cohort includes 145 children diagnosed with ADHD, 25 children with Autism Spectrum Disorder (ASD) who also meet the diagnostic criteria for ADHD, and 170 neurotypical controls (NC). In our analysis we used 120 ADHD children after excluding those with the ASD diagnosis, and matched this cohort with another 120 NC subjects. The data was downloaded from the Github repositories that are provided in [26].  
The acquisition protocol for the rs-fMRI data used a single shot, partially parallel gradient-recalled EPI sequence with TR/TE 2500/30ms, flip angle 70 degrees, and voxel resolution $3.05 \times 3.15 \times 3 mm^3$ on a Philips 3T Achieva scanner. The scan duration was either 128 or 156 time samples. Children were instructed to relax with their eyes open and focus on a central cross-hair, while remaining still for the scan duration. All participants completed a mock scanning session to habituate to the MRI environment. The rsfMRI data was pre-processed using a pipeline developed by CNIR. 
with 90 cortical/subcortical regions and 26 cerebellar regions, (2) the Harvard- Oxford atlas 
with 110 cerebral and cerebellar regions, and (3) the Craddock 200 atlas 
with 200 regions. The choice of different atlases enabled  analysis at multiple spatial scales. Additional demographic variables including age and sex, were also included in the analysis for both datasets.

\subsection{Classification Using Static Networks}
Here, we develop a novel Bayesian SVM classification approach based on their static brain network and supplemental covariates. Towards this aim, we considered a generic dataset with $N$ samples where the binary outcome $z_i \in \{-1, 1 \}$ depends on the vector of edge strengths $\bf {x}_i$ (of length $V(V-1)/2$), and additional non-network related covariate information as $\bf c_i$ $(C \times 1)$ for subject $i$.

\subsubsection{Static Functional Connectivity Estimation}
Assume $T_i$ scans for the $i$-th subject and denote the corresponding $V\times T_i$ fMRI data matrix as $\bm{Y}^{(i)}$, which contains time-series data from $V$ regions of interest (ROI). In order to construct the binary network that is used as a predictor in our SVM model, we use existing approaches in literature to estimate a sparse inverse covariance or precision matrix. In particular, we first calculated the Pearson correlation matrix (denoted by $\Theta_i$) for each subject using the observed fMRI data by averaging over all time points, and subsequently applied the graphical lasso algorithm [27] to compute a sparse inverse covariance matrix separately for each sample, although other algorithms can also be used. 
The off-diagonal elements in the sparse precision matrix $\Omega_i$ corresponding to the $i$th sample denotes the strength of the network edges in terms of the conditional dependency between two nodes in the network given all other nodes. For example, a zero off-diagonal element implies conditional independence between the corresponding nodes. The amount of structural zeros or the sparsity level of the network is controlled via a tuning or penalty parameter ($\lambda_{gl}$), with a larger $\lambda_{gl}$ value resulting in sparser networks. We describe a method to choose a suitable $\lambda_{gl}$ value in the sequel. 

{\noindent \underline{Preliminary edge screening:}} We propose a model-free approach for initial screening.  We screen out those edges with minimal variations in edge strengths across samples in order to exclude edges that are not instrumental in explaining between-subject differences from the analysis. For example, those edges that are present/absent in all or almost all samples are not instrumental for differentiating between groups of individuals and will be screened out. Such a screening step is able to reduce the dimension of the explanatory variables to be included in the model without compromising accuracy, while speeding up computations. We exclude all edges for which the variability in edge strengths across all sample was limited and did not exceed a certain threshold (standard deviation $\le 0.01$). More stricter thresholds were also considered but without any distinct improvements. We will denote the set of edges retained after screening as ${\bf x}^*_i$ ($Q\times 1$) that will be used in our model. We note that although the set of screened edges represent a viable set of candidate network connections that may drive differences across subgroups or classes, the final set of significant edges related to differences in class labels is inferred from amongst this set of initial screened edges under the fully Bayesian SVM approach, as elaborated in the sequel.

\subsubsection{Bayesian SVM via Pseudo-Likelihood}
We will concatenate the edge set and supplementary covariates to form an augmented feature vector denoted as $\bf u_i = [\bf x^*_i;c_i]$. 
We propose a linear classifier using a linear function of the features, i.e. $f(\bf{u_i})=\bm{u_i}^{T}  {\bm \beta}^* = {\bf x_i}^{*T} {\bm \beta} + {\bf c_i}^T {\bm\gamma}$. Here $\bm{\beta}$ is a $Q \times 1$ vector of edge-specific effects, ${\bm\gamma}$ is $C\times 1$ vector of non-network covariates, and ${\bm \beta}^*$ is of dimension $P\times 1$, where $P=Q+C$. Using the derivations in [28], the Bayesian SVM pseudo-likelihood can be represented as follows
\begin{eqnarray}
 &&   L(\bm{z} | \bm{u},\bm{\beta^*},\bm{\gamma},\sigma_{\epsilon}^2) 
    = \prod_{i=1}^N \frac{1}{\sigma_{\epsilon}^2}e^{-\frac{2}{\sigma_{\epsilon}^2} \max(1-z_i \bm{u_i}^T \bm{\beta^*}, 0)} \nonumber \\
&& =  \prod_{i=1}^N \int_{0}^{\infty} \frac{1}{\sigma_{\epsilon}\sqrt{2\pi \rho_i }}e^{-\frac{(\rho_i + 1 - z_i \bm{u_i^T \beta^*} )^2}{2\rho_i \sigma_{\epsilon}^2}}d\rho_i, \label{eq:lik}
\end{eqnarray}  
where, $\sigma_{\epsilon}^2$ represents the scale parameter in the likelihood, and a latent parameter $\rho$ is introduced to facilitate posterior computation. We note that the pseudo-likelihood is directly related to the hinge loss $\max(1-z_i \bm{u_i}^T \bm{\beta^*},0)$ that is routinely used in SVM models, where smaller value of the loss function implies large values of the pseudo-likelihood. Moreover, large values of $\sigma_{\epsilon}^2$ results in sharper changes in the pseudo-likelihood with changes in the values of the linear mean function. 




\subsubsection{Non-parametric Priors for Sparse Learning}
In practice, one expects only a subset of edges to be instrumental for classifying the binary outcome variable. Further, different edges may have different degrees of importance in the model. In order to ensure appropriate shrinkage of the edge effect sizes that allows for differential sparsity levels, we specify a Laplace or Double exponential (DE) prior on $\bm{\beta^*}=(\beta^*_1,\ldots,\beta^*_P)^T$ as: 
\begin{align}
\pi(\bm{\beta^*}) &= \prod_{p=1}^P \frac{\lambda_p}{2\sigma_{\epsilon}} e^{ -\frac{\lambda_p}{\sigma_{\epsilon}}|\beta^*_p|}, 
\lambda_p^2 \sim P, P\sim DP(M f_0),  \label{eq:BNP-DE} 
\end{align}
which is modulated by the local shrinkage parameter $\lambda_p>0$, that follows a non-parametric DP prior with precision parameter $M$ that controls the prior number of clusters (larger $M$ implies more clusters), and base measure $f_0 := Gamma(r,\delta)$ for ease in posterior calculations. Under a Bayesian specification, the prior on the scale parameter is specified as an inverse-Gamma prior with shape and scale parameters being $a_1,b_1,$ i.e. $\sigma_{\epsilon}\sim IG(a_1,b_1)$. The overall specification results in a novel DP mixture of Laplace (DPL) prior on ${\bm \beta}^*$, and we denote the corresponding approach as DPL-SVM. 

\begin{figure}[!h]
\centerline{\includegraphics[width=\columnwidth]{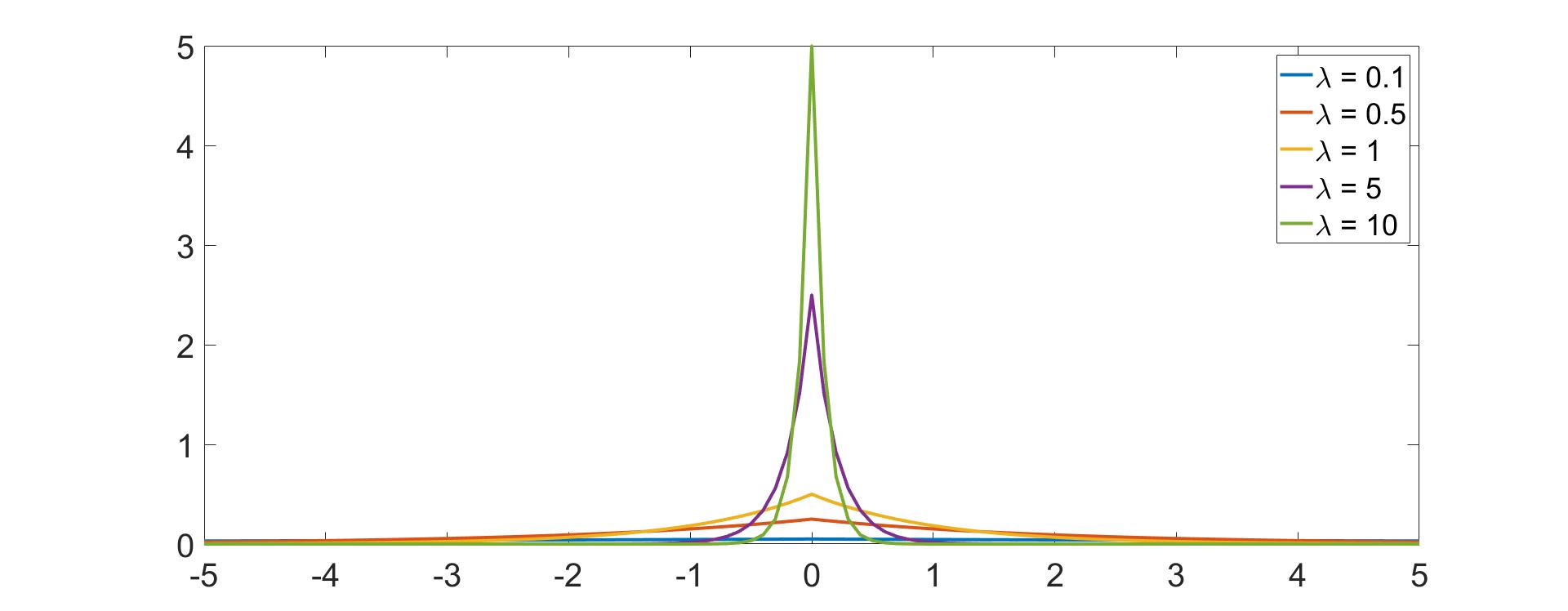}}
\caption{Plot of Laplace distribution under different values of the shrinkage parameter. X-axis represents different value of $\beta_{p}^*$. }
\label{fig:DEprior}
\end{figure}

The proposed approach is founded on the Bayesian Lasso method proposed by [29], where a large value of $\lambda_p$ implies greater prior shrinkage towards zero for $\beta_p$ (see Figure \ref{fig:DEprior}). However unlike in [29] that used global shrinkage, our approach is distinct in allowing for feature-specific shrinkage parameters. The novel specification in (\ref{eq:BNP-DE}) results in clustering of shrinkage parameters ($\lambda's$) across edges, where each cluster represents a distinct degree of shrinkage or importance of the coefficients in the SVM model. Moreover, the number of clusters and the cluster memberships are unknown and learnt in an unsupervised manner. The non-parametric prior defines a class of prior distributions on the set of densities for $\lambda$, where $f_0$ denotes the prior guess of the density. 
The following Lemma captures the closed form of the non-parametric mixture prior on the coefficients.

{\bf Lemma 1:} {\it The prior on the coefficients is given as $f(\beta^*_p) = \int \frac{\lambda}{2}\exp\{-\lambda |\beta^*_p| \}dP(\lambda) = \sum_{l=1}^\infty \pi_h f_h(\beta^*_p;\lambda^*_h)$, where $f_h(\beta^*_p;\lambda^*_h)=\frac{\lambda^*_h}{2}\exp\{-\lambda^*_h |\beta^*_p| \}$, $p=1,\ldots,P$, and $\pi_h = \nu_h\prod_{h'=1}^{h-1}(1-\nu_{h'}), \nu_h\sim Be(1,M)$, and $\lambda^*_h\stackrel{iid}{\sim} f_0$.}\\

Lemma 1 summarizes the fact the the prior distribution can be represented as an infinite mixture distribution with stick-breaking weights, where the shrinkage parameter for each component is drawn independently and identically (iid) from the base measure, i.e. $\lambda^*_h\stackrel{iid}{\sim}f_0$, which follows directly from the results in Sethuraman (1994). The infinite mixture representation highlights the non-parametric nature of the proposed prior. Further, using results in [30], the number of distinct clusters $\Delta$ for the shrinkage parameters $\lambda$ is guaranteed to increase with $P$ as well as the parameter $M$, under the rule $E(\Delta)=\sum_{m=1}^P M/(M+m-1)\approx M\log((P+M)/M)$, where $E(\Delta)$ represents the prior expectation of $\Delta$.

\begin{figure}[!h]
\centerline{\includegraphics[width=\columnwidth]{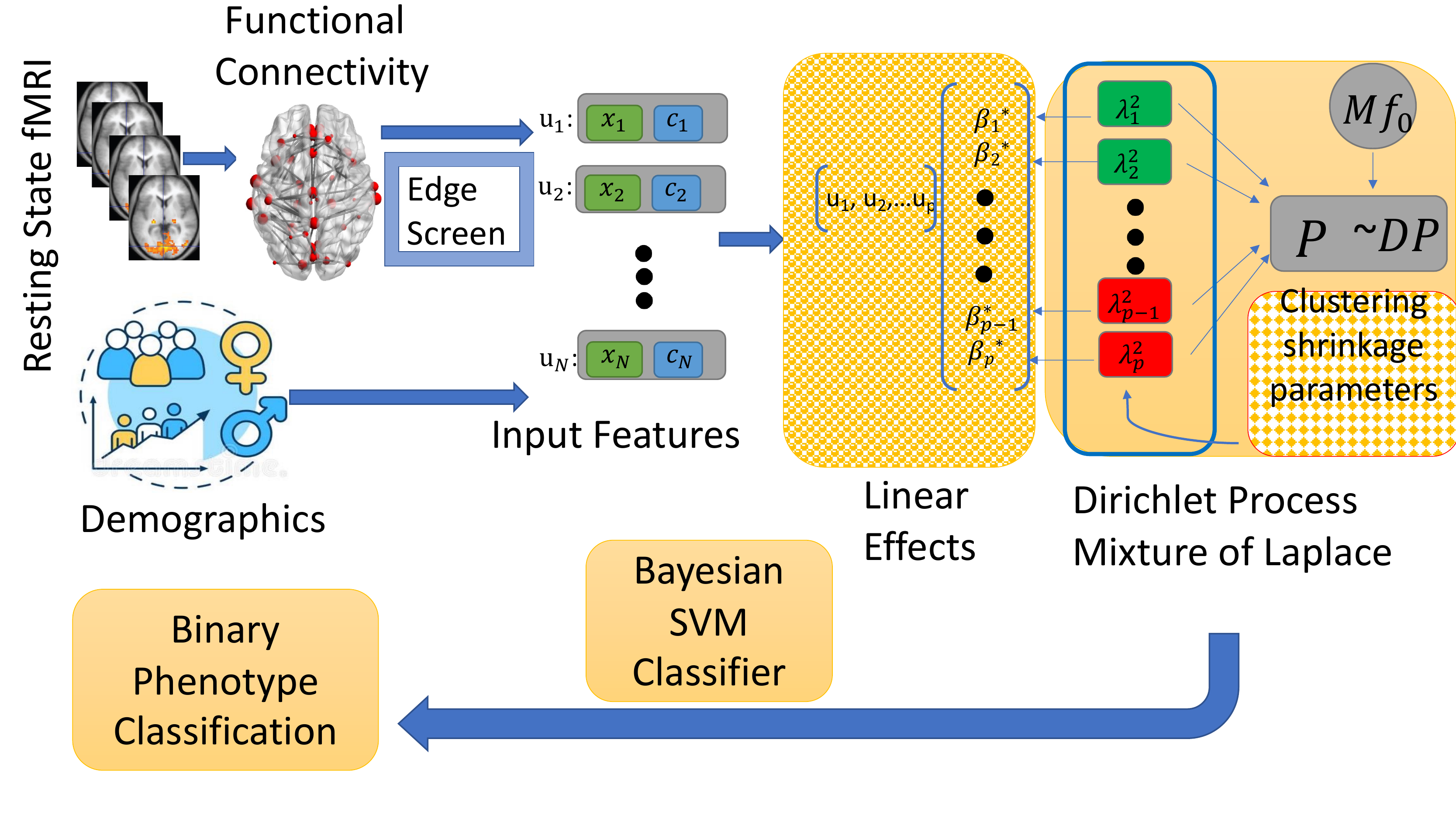}}
\caption{Illustration of the Bayesian SVM classifier based on static networks, using Dirichlet process mixture of Laplace priors on the feature effects. The resulting model leads to clustering of the feature-specific shrinkage parameters, with each cluster represented by a distinct color, resulting in an unsupervised method to learn differential important levels of the features by pooling information across network edges.}
\label{fig:schema}
\end{figure}

The proposed approach results in sharp improvements over parametric approaches that are not necessarily equipped to combine information across edges, by learning differential sparsity levels (represented by $\lambda$) across subsets of edges in an unsupervised manner. 
For example, the traditional Bayesian lasso approach [29] specifies a global shrinkage parameter $\lambda$ across all edges, which is an extreme case of the proposed model with $\Delta=1$, but can not capture differential sparsity patterns across features. Another special case is when $\Delta=P$ that forces unique shrinkage parameters across all edges. However, this approach is expected to result in poor accuracy due to the inability to robustly estimate the importance weights by pooling information across edges. The proposed Bayesian SVM method (illustrated in Figure \ref{fig:schema}) provides a desirable balance between these two extreme scenarios, and results in data-adaptive learning of differential sparsity levels across unknown subsets of edges.

\subsubsection{Posterior Computation}
In order to facilitate posterior calculations, we re-express the Laplace distribution as a scale-mixture of normals with an exponential mixing density as $
    \beta^*_p | \sigma_{\beta_p}^2 \sim N(0, \sigma_{\beta_p}^2), \mbox{ } 
    \sigma_{\beta_p}^2 \sim \frac{\lambda_p^2}{2} exp\big(-\frac{\lambda_p^2}{2}\sigma_{\beta_p}^2 \big) $,
where $\lambda_p^2 \sim P, \mbox{ } P\sim DP(M f_0)$.
The model parameters are updated via a Markov Chain Monte Carlo sampling (MCMC) sampling scheme. A fully Gibbs sampler was used for posterior computation, which cycles through the following update steps.

{\noindent \bf Algorithm I: MCMC Steps for DPL-SVM }

{\noindent (i) Use the following conditional distribution to sample $\sigma_{\epsilon}^2$  -
$ IG\big(a_1 + \frac{3N}{2}, b_1 + \sum_{i=1}^N \frac{(\rho_i + 1 -z_i \bm{u_i}^T \bm{\beta^*}){^2}}{2\rho_i} \big) $.}

{\noindent (ii) $\rho_i$ is sampled using the inverse Gaussian conditional distribution} $ IN\big(|1 -z_i \bm{u_i}^T \bm{\beta^*} |^{-1}, \frac{1}{\sigma_{\epsilon}^2}\big), i=1,\ldots,N$.

{\noindent (iii) Use a multivariate normal distribution with mean $\bm \mu_{\beta}$ and variance $\bm \Sigma_{\beta}$ to sample $\bm \beta^*$, where 
   $\bm \Sigma_{\beta} = \big(\frac{1}{\sigma_{\epsilon}^2} \sum_{i=1}^N {\bf u}_i{\bf u}^T_i + D^{-1}_{\sigma_\beta} \big)^{-1} , \mbox{ } 
    \bm \mu_{\beta} = \Sigma^{-1}_{\beta}\{ \frac{1}{\sigma^2_\epsilon}\sum_{i=1}^N \rho^{-1}_i z_i\bm{u}_i (\rho_i + 1 )$ \}, and $D_{\sigma_\beta}$ represents a diagonal matrix with entries  $(\sigma_{\beta_1}^2 \dots \sigma_{\beta_P}^2)$. } 

{\noindent (iv) Use  $\pi(\sigma_{\beta_p}^{-2} | -) = IN(\big | \frac{\lambda_p}{\beta^{*}_{p}}\big |,\lambda_p^2)$ to update $\sigma_{\beta_p}^2$.}


{\noindent (v) The cluster memberships under the DP and the stick-breaking weights $\nu$ are updated under the slice sampler [31] and not presented here due to space restrictions. Conditional on the cluster memberships, the shrinkage parameter corresponding to the $h$-th cluster is updated as: $Ga\big(\lambda^*_h; n_{h}+ r, \delta + 0.5\sum_{j=1}^P 1(H_j=h) |\beta^*_j| \big)$, where $H_j$ denotes the cluster membership for $j$th shrinkage parameter, $1(\cdot)$ is the indicator function, and $n_h=\sum_{j=1}^p 1(H_j=h)$}.

\subsubsection{Choice of the network sparsity:} The proposed DPL-SVM approach may be sensitive to the choice of the estimated static network that is used as the input feature vector for classification. In order to select the optimal network sparsity level, we fit the graphical lasso algorithm over a grid of shrinkage parameter values, and fit the proposed DPL-SVM based on the corresponding edge set from the estimated static network. For a given network, we calculate the classification accuracy in a validation sample. Subsequently, we choose the  network sparsity level that results in the best validation accuracy, and use this choice for out-of-sample classification for test sample. Such a validation scheme provides a key understanding of network sparsity levels yielding optimal classification.

\subsection{Extension to Classification Using Dynamic Networks}
It is of interest to evaluate how the resting state static FC compares with resting state dynamic FC in terms of classification accuracy. To this end, we extend the Bayesian SVM to include dynamic connectivity features. Dynamic FC involves a time-varying connections for each edge, which need to be integrated into the model, thus posing methodological challenges.  Instead of using the full time-series, we use a set of extracted features from the connectivity time-series for each edge with minimal loss of information. These methods included (i) the three manually specified features used in [19] that involved the mean FC along with measures of variation and stability of the time-varying FC; (ii) features extracted from a principal component analysis (PCA) of the time-varying functional connectivity, that retained at least 95\% variability; and (iii) features extracted from independent component analysis (ICA), where the number of components was chosen to be the same as in the PCA analysis. Hence, our analysis is able to evaluate which set of extracted features yields the best classification performance. 

{\noindent \underline{Dynamic Connectivity Estimation:}} 
We use the sliding window approach for computing dynamic FC  [32], that results in a time-series of pairwise correlation matrices $\tilde{\Sigma}_W,\ldots,\tilde{\Sigma}_{T-W}$ capturing connectivity strengths, where $W$ denotes the window length that is determined via cross-validation. 
We extract summary features for each edge that reduces dimensionality with minimal loss of information. In particular, consider the extracted features $\{u_{kl,1},\dots,u_{kl,R}\}$ from the time-series of sliding window correlations $\{\tilde{\sigma}_{kl,2},\ldots,\tilde{\sigma}_{kl,T-1} \}$ corresponding to edge $(k,l)$, where $R$ is pre-specified. For subject $i$, let $\bm{u_i^r}$ ($Q\times 1$) be denote the subset of screened edges for feature $r,r = 1\dots R$.



\subsubsection{Proposed Model}

 The Bayesian SVM pseudo-likelihood based on dynamic FC may be represented as $L(\bm{z} | \bm{u},\bm{\beta^*},\bm{\gamma},\sigma_{\epsilon}^2) 
    = \prod_{i=1}^N \int_{0}^{\infty} \frac{1}{\sigma_{\epsilon}\sqrt{2\pi \rho_i }}exp\big(-\frac{(\rho_i + 1 - z_i\sum_{r=1}^R \bm{u_i}^{rT} \bm{\beta_r^*} - {\bf c_i}^T\bm{\gamma})^2}{2\rho_i \sigma_{\epsilon}^2}\big)d\rho_i $. 
Noting that the effects for the extracted features corresponding to the same edge are expected to be correlated, we propose a structured regression coefficient as $\beta^*_{q,r} = \beta^*_{q} \eta_r, r=1\dots R, 1\le q \le Q$, where $\{\eta_1,\ldots,\eta_R \}$ represent feature-specific correlated effects that are considered invariant across edges that is modeled as $\bm{\eta} \sim N(\bm{0},\bm{\Sigma_{\eta}})$, with $\bm{\Sigma_{\eta}} \sim IW(b,D),b=R,D=d^* \times I_R $, $d^* \sim IG(c,d)$, such that $\bm{\Sigma_{\eta}}$ captures the correlations within $\bm{\eta}$ and $IW(\cdot)$ denotes the inverse-Wishart prior. On the other hand, $\beta^*_{q}$ reflects the global edge-specific contribution of the extracted features in the SVM model, with $\beta^*_{q}=0$ implying no effect corresponding to the dynamic connection of edge $q$. The edge-specific parameters $(\beta^*_1,\ldots,\beta^*_Q)$ are assigned the same DP mixtures of Laplace prior as in (\ref{eq:BNP-DE}). 



We note that while the structured coefficients in the dyDPL-SVM model  $\beta_{q,r} = \beta^*_{q} \eta_r (r=1\dots R, 1\le q \le Q)$ can be generalized to accommodate greater flexibility, such modeling assumptions can lead to a dramatic increase in the number of parameters and slow down computations. For example, one can instead specify $\beta_{q,r} = \beta^*_{q} \eta_{q,r}$, which allows for edge-specific and feature-specific variations. However, it is clear that the overall number of parameters for such a specification becomes exceedingly large with the increase in the network size, resulting in computational bottlenecks and potential overfitting issues. Hence we do not consider such generalizations further.  

\subsubsection{Posterior Computation}
The form of the pseudo-likelihood using dynamic connectivity features lends itself to a similar treatment for MCMC sampling as in model (\ref{eq:lik}) with static connectivity features, with some additional steps needed to sample the $\eta$ effects and associated hyperparameters.


{\bf \noindent Algorithm II: MCMC Steps based for dynamic DPL-SVM} \\
The posterior distributions of $\sigma_{\epsilon}^2$ and $\rho_i^{-1}$ are given similarly as in steps (i) and (ii) in Algorithm I involving static connectivity, but with the linear term in the exponent now modified to be $z_i\sum_{r=1}^R \bm{u_i}^{rT} \bm{\beta_r^*} + {\bf c_i}^T\bm{\gamma}$. Similarly the updates for  $\sigma_{\beta_p}^2$ proceeds similarly as in Algorithm I. Moreover, the posterior distributions for $\bm \beta^*$ follows multivariate normal distribution with mean $\bm \mu_{\beta}$ and variance $\bm \Sigma_{\beta}$, where 
  $\bm \Sigma_{\beta} = \big(\frac{1}{\sigma_{\epsilon}^2} \sum_{i=1}^N {\bf u}_{\eta,i}{\bf u}^T_{\eta,i} + D^{-1}_{\sigma_\beta} \big)^{-1} , \mbox{ } 
    \bm \mu_{\beta} = \Sigma^{-1}_{\beta}\{ \frac{1}{\sigma^2_\epsilon}\sum_{i=1}^N \rho^{-1}_i z_i\bm{u}_{\eta,i} (\rho_i + 1  - {\bf c_i}^T\bm{\gamma} )$ \}, where $ {\bf u}_{\eta,i}=\sum_{r=1}^R\bm{u_i}^{r}\eta_r$ and $D_{\sigma_\beta}$ represents a diagonal matrix with entries  $(\sigma_{\beta_1}^2 \dots \sigma_{\beta_P}^2)$.  Similarly,  the posterior distributions of $\bm{\eta}$ follows the Multivariate Normal distribution with mean $\bm{\mu_{\eta}}$ and variance $\bm{\Sigma^*_{\eta}}$, with 
   $ \bm{\Sigma^*_{\eta}} =\sum_{i=1}^N \bigg \{ \widehat{\bm{U_{\eta,i}}}^T \bm{\beta^*} \bm{\beta^{*T}}\widehat{\bm{U_{\eta,i}}} \bigg \} + \bm{\Sigma}_{\eta}, \mbox{ }
    \bm{\mu_{\eta}} = \sum_{i=1}^N \bigg \{\frac{\bm{\beta^{*T}}\widehat{\bm{U_{\eta,i}}} (\rho_i + 1-{\bf c_i}^T\bm{\gamma})}{\rho_i \sigma_{\epsilon}^2}   \bigg \}  \bm{\Sigma^*_{\eta}}$,
where  $\widehat{\bm{U_{\eta,i}}}$ is $R\times Q$ matrix with the $r$th row as $\bm{u_{ir}}, r=1\dots R$. The posterior distribution for updating $\bm{\Sigma}_{\eta}$ is given as 
$\bm{\Sigma_{\eta}} \sim IW(b+ 1,D+\bm{\eta}^T \bm{\eta})$ 
The conditional distribution of $\lambda_p^2$ is given similarly as that in the posterior update in step (v) in Algorithm I, and the ${\bm \gamma}$ effects are updated similarly as ${\bm \beta}^*$.

\subsubsection{Choice of window length} 
In our approach focused on classification, we fit the proposed dyDPL-SVM model separately over a grid of window length values, and compute the classification performance in a validation sample. Subsequently, we choose the window length that results in the lowest mis-classification accuracy in the validation sample. 

\subsection{Classification via Integrating Multiple fMRI Sessions}
One major question of interest in the neuroimaging community, is whether it is possible to develop classification approaches with higher accuracy via integrating information across multiple sessions. To address this question, we used static networks computed using resting state fMRI data from LR1 and LR2 sessions together in the proposed approach based on static connectivity. In other words, the edge sets for both the LR1 and LR2 networks were included jointly in the Bayesian SVM model for classification, where the network density for these networks was determined from the single-session static connectivity analysis as in Section B.5. A preliminary edge screening was performed as in Section B.1 for each of the two scans, and subsequently the union of these two subsets was used for classification.

\subsection{Feature Selection}
The fully Bayesian framework seamlessly enables one to perform inference. For example, testing for significant coefficients is by constructing $100(1-\alpha)\%$ credible intervals based on the posterior distributions of $\beta_1,\ldots,\beta_Q$. Here $\alpha$ is the level of the credible set and it can be adjusted to account for multiplicity corrections. The resulting approach enables one to identify the significant network edges and other covariates that play an important role in classification under the SVM model.
In order to determine which edges are consistently significant in the SVM model, we randomly divide the overall sample into multiple training and validation slices in the ratio 90\%-10\% and then perform feature selection in the training sample for each split. Subsequently, we identify those network edges that show up as significant across all the multiple random splits as reproducible features that significantly influence classification. 
Given that we are able to achieve almost perfect classification for the top and bottom 10\% fluid intelligence groups corresponding to the LR1 scan of the HCP study (see results in the sequel), we focus on this subset of individuals for feature selection analysis.
The reproducibility analysis is not performed for the CNI data given that the classification performance is not as optimal as in the HCP analysis; instead we simply report the significant edges identified under the full analysis with all CNI data samples included. 

\section{Numerical Results Using rs-fMRI Data}
\subsection{Study Objectives}
\subsubsection{Human Connectome Project Data} Our goal is to use resting state functional connectivity to classify individuals belonging to various strata in the intelligence spectrum. To this end, we consider classifying individuals belonging to the top and bottom $\zeta\%$ intelligence groups, where $\zeta$ is chosen to be $10\%, 12\%, 15\%, 18\%,$ to reflect different levels of intelligence. The sub-population from the top and bottom $18\%$ of the population that was used for fluid intelligence/crystallized intelligence analysis had a mean age of 28.7/28.9 with 47.4\%/48.5\% males in the top 18\% group and 44.7\%/39.4\% males in the bottom 18\% group. The mean (s.d.) of fluid intelligence for the top and bottom 18\% groups were 129.1(9.5) and 99.8(3.9) respectively, and they were 121.2(7.7) and 103.6(4.2) respectively for crystallized intelligence. Similarly, the mean (s.d.) of fluid intelligence for the top and bottom 10\% groups were 136.6(4.0) and 97.1(3.2) respectively, and they were 134.8(5.1) and 99.6(3.5) respectively for crystallized intelligence.
We perform the analysis separately for both fluid intelligence and crystallized intelligence, with the goal being to classify them into high or low intelligence groups. We note that while the classification accuracy is expected to be high when classifying the top and bottom 10\% intelligence levels, it is likely to deteriorate for larger sub-populations with lesser separation between the low and high intelligence groups. Age and gender are included as explanatory variables, along with the the brain network calculated under the Glasser atlas with 360 nodes. We use rs-fMRI data from both LR1 and LR2 scans for analysis.

We perform three different streams of analysis for each intelligence strata denoted by $\zeta$. First we use static FC to evaluate classification performance using data from one scanning session, separately for the LR1 and LR2 scans from HCP. Subsequently, we repeat this analysis with dynamic connectivity for these scans. Finally, we combine the resting state static networks from LR1 and LR2 scans to perform an integrative classification analysis. The three separate analysis provides us with an understanding of which types of network-based features (single-session or multi-session analysis based on static networks, or dynamic networks) derived from resting-state fMRI data is expected to yield superior classification accuracy. For the static connectivity analysis using one scanning session, we computed the network across varying sparsity levels by tuning the $\lambda_{gl}$ parameter over a grid for each individual, and we chose the network density for all individuals that corresponded to a pre-determined sparsity level. We considered a range of network sparsity values varying from 12\% to 25\%, and use the previously described pre-screening step to select a subset of edges to be used in the classifier for each level of network density. 
For the multi-session analysis, we used the network sparsity level that was deemed  optimal under the single session analysis for LR1 and LR2 scans. 

\subsubsection{ADHD Data}
In this analysis, we used resting state fMRI data from the CNI-TLC data, to classify ADHD individuals versus controls based on resting state FC. 
 The mean (s.d.) age for the ADHD and TC groups were 10.4(1.5) and 10.3(1.2) respectively, and the gender distribution for these groups comprised 31.0\% boys for ADHD and 30.0\% boys for TC. The SVM model was trained separately on data coming from the three parcellations AAL (116 ROIs), Harvard Oxford (110 ROIs) and Craddock200 (200 ROIs), and the relative performance over the three parcellations was compared. Coupled with the HCP analysis, this second analysis highlights the generalizabilty of the proposed DPL-SVM approach for different disease areas, and different choices of brain parcellations. 

\subsection{Comparison Methods and Metrics}
We evaluate the performance of the proposed method in the context of several state-of-the-art competing methods for classification. The first competing method is penalized SVM 
that uses a combination of SCAD and ridge penalties for estimating the coefficient effects to overcome the limitations of each penalty applied separately. The second competing method is penalized logistic regression, where the elastic net penalty (the combination of L-1 and L-2 penalties) was used. As the performance of the penalized methods (denoted as Pen-SVM and Pen-Logit and implemented via the R packages `penalizedSVM' and `penalized' respectively) are sensitive to the choice of tuning parameters, they were trained over a grid of tuning parameters and the cross-validation was used to select the best tuning parameter. The third competing method is the Bayesian SVM method that imposes the following prior: $\beta_p \stackrel{indep}{\sim} DE(\lambda_p), \lambda_p\stackrel{indep}{\sim} Ga(a_\lambda,b_\lambda), p=1,\ldots,P$. This approach (denoted as Ind-Bayes) is distinct from the proposed DP mixture of Laplace distribution in terms of not being able to cluster shrinkage parameters across edges to learn differential sparsity patterns. The last competing method considered is the Naive Bayes classifier (denoted as N-Bayes) that uses a Bayes rule for classification, and was implemented via the `naiveBayes' R package.   

We evaluate the performance of different approaches using out-of-sample classification accuracy of the  test samples, via several metrics popularly used in literature [26].  In particular, we report the mis-classification rate defined as the  ratio of incorrect classification labels over the total number of instances evaluated, the $F1$ score that is computed as the harmonic mean between recall (sensitivity) and precision, i.e. 2 recall$\times$ precision/(recall+precision), and the informedness that is also known as Youden's J statistic and summarizes the true positive and true negative rates as sensitivity + specificity -1. We note that precision is defined as the fraction of correctly classified positive samples in relation to the number of the total positive classified samples, while sensitivity is defined to be the fraction of positive labels that are correctly classified and specificity is the fraction of correctly classified negative labels. In general, superior performance is indicated via low mis-classification accuracy and higher $F1$ score, and Youden's J statistic. Throughout the presentation of results, bolded number were used to imply significantly improved results compared to all the other methods.

\subsection{Analysis Results for HCP Data}

\begin{figure}
    \centering
    \includegraphics[width=\linewidth]{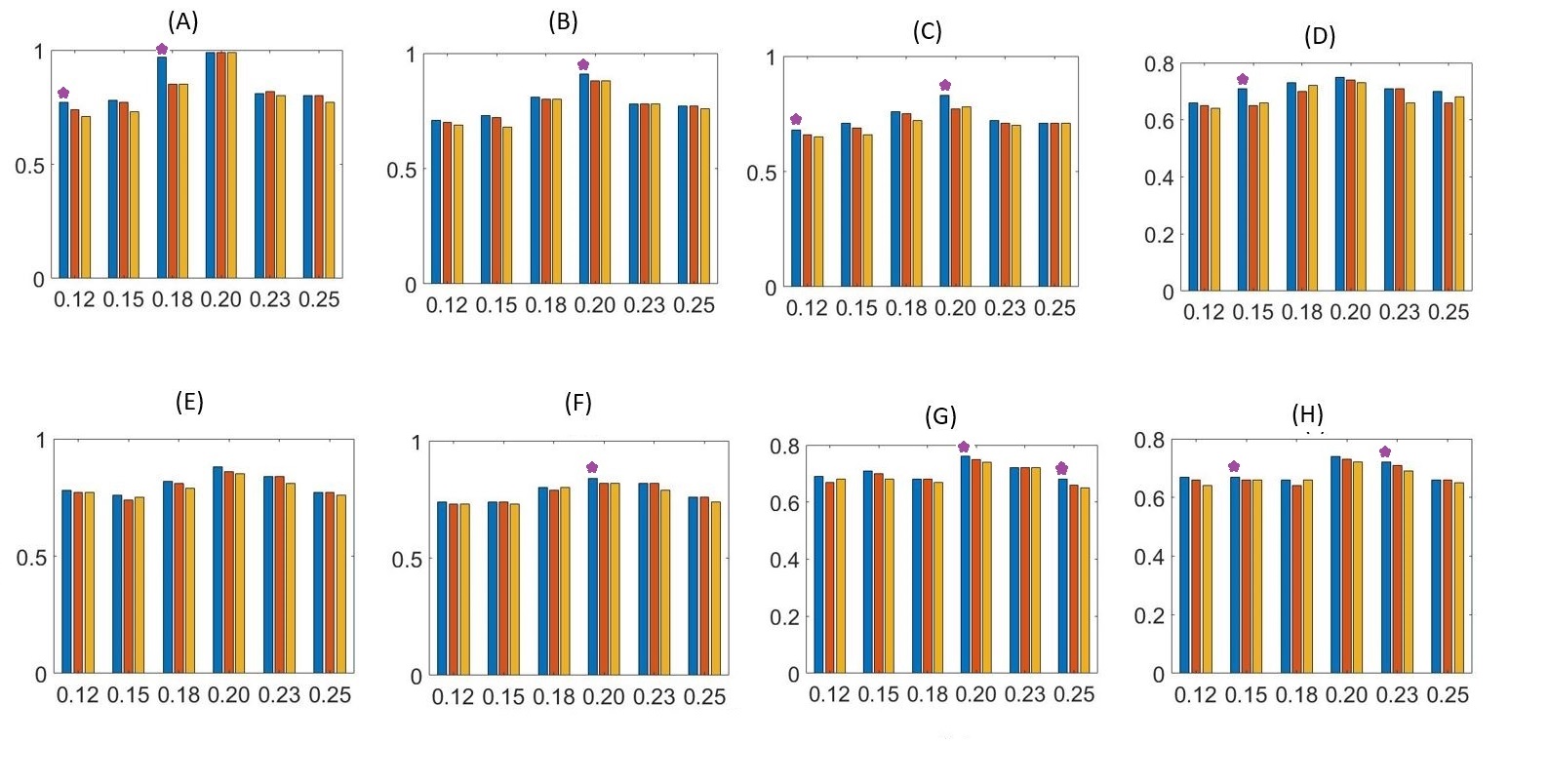}
    \caption{F1-score for intelligence classification using static functional connectivity based on HCP-LR1 data with different sub-populations and across varying network densities (X-axis). Colored bars represent the different approaches - Blue: DPL-SVM, Red: PenSVM, Yellow: PenLogit. The other methods were omitted due to consistently poor performance. Panels A-D and E-H pertain to sub-populations of fluid and crystallized intelligence respectively - A/E: 10\%, B/F: 12\%, C/G: 15\%, D/H: 18\%.  }
    \label{fig:F1-static}
\end{figure}

{\noindent\underline{Results based on Single-Session Static Connectivity:}} As seen from Figure \ref{fig:F1-static}, the proposed approach almost always resulted in superior classification accuracy across all the network sparsity levels compared to competing methods. Moreover, the best classification accuracy for both fluid and crystallized intelligence was obtained with network density as 0.2 under the LR1 scan, and 0.18 for the LR2 scan under the proposed approach. This finding is of independent interest, and can potentially shed light on the optimal network densities required to obtain good classification accuracy for intelligence based on static connectivity. The results for classification performance using static connectivity for the HCP data are reported in Table \ref{tab:static_results} under these optimal network sparsity levels for all methods, corresponding to both fluid intelligence and crystallized intelligence. The median number of clusters across all MCMC iterations under the DPL-SVM approach ranged from 2-4 for both fluid and crystallized intelligence classification. 

Several aspects are clear from the reported results. First, the classification accuracy under all approaches was the highest when classifying the top and bottom 10\% intelligence groups, but the accuracy deteriorates as the size of these subgroups are increased gradually. This is expected given the decreasing separation between high and low intelligence groups for more heterogeneous sub-populations. Second, the proposed DPL-SVM approach has statistically significant improvements for almost all cases, barring a few exceptions. Competing approaches consistently had inferior classification performance due to their inability to adaptively pool information across features in order to determine differential sparsity levels. These results clearly illustrate that the proposed  approach provides consistent and reliable improvements across varying intelligence spectrum, which highlights it's utility as a classification method for connectome fingerprinting. Third, by including age and gender along with the static network in the model, it is possible to obtain improved accuracy compared to using only network-based in the SVM model (results not reported here). This illustrates the importance of including demographic features (particularly age) in intelligence classification.


Fourth, we discover that the classification performance is generally improved under the LR1 scan compared to the LR2 scan under all methods for fluid intelligence classification. This is supported by the definition of fluid intelligence that measures intuitive and spontaneous reasoning, and given that it is reasonable to conjecture that more intuition was used for the first scan compared to the second scan when participants were already acclimated to the experience of fMRI scanning. On the other hand, classification accuracy for crystallized intelligence was often higher corresponding to the LR2 scan compared to LR1 under all approaches, which is consistent with the definitions of crystallized intelligence that is based on accumulated knowledge from the past. 

\begin{table}[]
    \centering
\scriptsize
\begin{tabular}{|l|lll|lll|lll|}
    \hline
  &\multicolumn{9}{c|}{Fluid Intelligence Classification}\\
  \hline
     &\multicolumn{3}{|c|}{LR1}&\multicolumn{3}{|c|}{LR2} &\multicolumn{3}{|c|}{Multi-session} \\
     $\zeta$=10\%  &MC&F1&I &MC&F1&I &MC&F1&I\\  
     \hline
     DPL-SVM &0.00&0.99&0.99 &0.05&\bf{0.97}&\bf{0.93} &0.00&0.99&0.99\\
     Pen-SVM &0.00&0.99&0.99&0.06&0.93&0.88 &0.00&0.99&0.99 \\
     Pen-Log & 0.00&0.99&0.99&0.06&0.93&0.88 &0.00&0.99&0.99 \\
     Ind-SVM &0.25 &0.74 &0.51 &0.25 &0.75 &0.51 &0.25&0.76&0.52\\
     N-Bayes &0.04 &0.96 &0.90 &0.18 &0.82 &0.64 &0.09&0.93&0.83\\
     \hline
    $\zeta$=12\%  &MC&F1&I &MC&F1&I &MC&F1&I\\   
    \hline
     DPL-SVM &\bf{0.09}&\bf{0.92}&\bf{0.84}&\bf{0.14}&\bf{0.87}&\bf{0.72} &0.07&0.94&0.87\\
     Pen-SVM &0.12&0.87&0.75&0.17&0.83&0.68 &0.07&0.93&0.87\\
     Pen-Log &0.12&0.87&0.75&0.18&0.81&0.63 &0.08&0.91&0.83\\
     Ind-SVM &0.27 &0.71 &0.46 &0.26 &0.73 &0.54 &0.24&0.77&0.55\\
     N-Bayes &0.25 &0.71 &0.56 &0.31 &0.76 &0.41 &0.06&0.95&0.88\\
     \hline
    $\zeta$=15\%  &MC&F1&I &MC&F1&I &MC&F1&I\\   
    \hline
     DPL-SVM &\bf{0.17}&\bf{0.85}&\bf{0.68}&0.25&\bf{0.78}&\bf{0.54} &0.14&\bf{0.87}&\bf{0.73}\\
     Pen-SVM &0.19&0.80&0.62&0.26&0.73&0.48 &0.15&0.84&0.70\\
     Pen-Log &0.21&0.80&0.61&0.27&0.72&0.46 &0.15&0.85&0.71\\
     Ind-SVM &0.45 &0.28 &0.11 &0.30 &0.70 &0.42 &0.26&0.74&0.50 \\
     N-Bayes &0.47 &0.15 &0.04 & 0.44&0.30 &0.10 &0.40&0.62&0.22 \\
     \hline
    $\zeta$=18\%  &MC&F1&I &MC&F1&I &MC&F1&I\\   
    \hline
    DPL-SVM  &\bf{0.24}&\bf{0.78}&\bf{0.52}&\bf{0.22}&\bf{0.78}&\bf{0.53} &\bf{0.22}&\bf{0.79}&\bf{0.56}\\
     Pen-SVM &0.26&0.73&0.48&0.26&0.73&0.47&0.25&0.74&0.50\\
     Pen-Log &0.27&0.72&0.46&0.27&0.71&0.43 &0.26&0.73&0.48\\
     Ind-SVM &0.32 &0.68 &0.36 &0.33 &0.67 &0.34 &0.32&0.68&0.37 \\
     N-Bayes &0.26 &0.73 &0.49 &0.24 &0.76 &0.50 &0.24&0.76&0.51 \\
     \hline
     \hline
        &\multicolumn{9}{c|}{Crystallized Intelligence Classification}\\
          \hline
     $\zeta$=10\%  &MC&F1&I &MC&F1&I &MC&F1&I\\    
     \hline
     DPL-SVM &0.13&\bf{0.88}&\bf{0.75}&0.11&\bf{0.91}&{\bf 0.81} &0.09&0.92&0.84\\
     Pen-SVM &0.14&0.86&0.73&0.11&0.89&0.79 &0.09&0.91&0.83 \\
     Pen-Log &0.15&0.85&0.70&0.12&0.88&0.75 &0.10&0.91&0.82 \\
     Ind-SVM &0.31 &0.69 &0.39 &0.30 &0.70 &0.42 &0.30&0.71&0.42\\
     N-Bayes &0.22&0.76&0.56&0.15 &0.84 &0.71 &0.15 &0.85 &0.72  \\
     \hline
     $\zeta$=12\%   &MC&F1&I &MC&F1&I &MC&F1&I\\   
     \hline
     DPL-SVM &0.17&\bf{0.85}&0.68&0.14&\bf{0.86}&{\bf 0.72} &0.12&\bf{0.89}&\bf{0.77}\\
     Pen-SVM &0.18&0.82&0.68&0.15&0.84&0.70 &0.13&0.86&0.72\\
     Pen-Log &0.18&0.81&0.64&0.17&0.82&0.67 &0.14&0.84&0.70\\
     Ind-SVM &0.32 &0.68 &0.36 &0.30 &0.70 &0.40 &0.30&0.71&0.41\\
     N-Bayes &0.23&0.79&0.49&0.23 &0.71 &0.56 & 0.23&0.80 &0.56 \\
     \hline
    $\zeta$=18\%  &MC&F1&I &MC&F1&I &MC&F1&I\\  
    \hline
     DPL-SVM &\bf{0.25}&\bf{0.77}&\bf{0.53}&0.25&\bf{0.77}&{\bf 0.51} &\bf{0.22}&\bf{0.77}&\bf{0.56}\\
     Pen-SVM &0.27&0.73&0.49&0.25&0.74&0.49  &0.24&0.74&0.50\\
     Pen-Log &0.28&0.73&0.48&0.27&0.71&0.44 &0.26&0.73&0.48\\
     Ind-SVM &0.33 &0.66 &0.33 &0.34 &0.65 &0.32 &0.33&0.67&0.34\\
     N-Bayes &0.41&0.46&0.12&0.41 &0.39 &0.25 &0.41 &0.50 &0.12 \\
     \hline
    \end{tabular}
    \caption{Classification results for HCP data based on static connectivity, using single- and multi-session analysis. Results are reported under varying sub-populations corresponding to the top and bottom $\zeta\%$ of intelligence spectrum. MC, F1, and I refer to the mis-classification rate, the F1 score and the informedness metrics. }
    \label{tab:static_results}
\end{table}


\begin{table}[h!]
      \centering
          \scriptsize
    \begin{tabular}{|l|l|ccc|ccc|}
    \hline
  Extracted  & Fluid   &MC&F1&I &MC&F1&I\\
  Features &Intelligence   & \multicolumn{3}{c|}{HCP-LR1}  & \multicolumn{3}{c|}{HCP-LR2}  \\    
     \hline
 Sliding  &DPL-SVM  & 0.06 &\bf{0.95}& \bf{0.89} &0.20&0.80&0.60 \\
  Window   & Pen-SVM  & 0.07 &0.93&0.87&0.21&0.79&0.59 \\
     &Pen-Log &0.07&0.92&0.86&0.22&0.78&0.57\\
     & Ind-SVM &0.30&0.71 &0.40 &0.33 &0.67 &0.36 \\
     &N-Bayes &0.30 &0.71 &0.41 &0.26 &0.73 &0.48\\
          \hline
PCA    &  DPL-SVM & 0.36 &\bf{0.68}&\bf{0.33}&\bf{0.36}&\bf{0.67}&\bf{0.31}\\
&     Pen-SVM &0.37 &0.64 &0.25&0.39&0.63&0.25\\
&     Pen-Log &0.42&0.58&0.19&0.42&0.58&0.19\\
&     Ind-SVM &0.39&0.61 &0.23 &0.39 &0.60 &0.22\\
&     N-Bayes &0.44 &0.37 &0.12 &0.35 &0.61 &0.28\\
     \hline
 ICA  &    DPL-SVM & 0.33 &0.71&0.39&0.35&\bf{0.67}&\bf{0.32}\\
 &    Pen-SVM &0.35 &0.65 &0.30&0.36&0.63&0.24\\
 &    Pen-Log &0.41&0.60&0.22&0.43&0.57&0.18\\
 &     Ind-SVM &0.40 &0.60 &0.22 &0.38 &0.63 &0.25\\
 &    N-Bayes &{\bf0.26} &0.70 &0.40 &0.41 &0.40 &0.10\\
     \hline
    \hline 
Extracted &   Crystallized  &MC&F1&I &MC&F1&I\\
    
Features  &Intelligence    & \multicolumn{3}{c|}{HCP-LR1}  & \multicolumn{3}{c|}{HCP-LR2}  \\    
     \hline
  Sliding   &    DPL-SVM  & 0.18&\bf{0.82}&\bf{0.69}&\bf{0.16}&\bf{0.86}&\bf{0.73}\\
 Window &    Pen-SVM  & 0.19&0.80&0.66&0.18&0.82&0.69\\
 &    Pen-Log  & 0.19&0.80&0.65&0.19&0.80&0.62\\
 &     Ind-SVM & 0.37&0.64 &0.27 &0.34 &0.66 &0.32\\
 &    N-Bayes &0.37 &0.50 &0.14 &0.31 &0.60 &0.22\\
          \hline
 PCA&    DPL-SVM & 0.34 &0.66&0.33&0.30&\bf{0.71}&\bf{0.49}\\
 &    Pen-SVM &0.35 &0.63 &0.31&0.31&0.68&0.40\\
 &    Pen-Log &0.35&0.62&0.30&0.32&0.66&0.36\\
 &     Ind-SVM &0.39 &0.61 &0.23 &0.37 &0.64 &0.26\\
 &    N-Bayes &{\bf 0.23} &0.66 &{\bf 0.40} &0.44 &0.35 &0.15\\
     \hline
 ICA &    DPL-SVM & \bf{0.30} &\bf{0.72}&\bf{0.48}&0.30&\bf{0.71}&\bf{0.48}\\
 &    Pen-SVM &0.32 &0.67 &0.36&0.31&0.69&0.44\\
 &    Pen-Log &0.34&0.60&0.30&0.32&0.66&0.36\\
 &     Ind-SVM &0.38 &0.61 &0.23 &0.37 &0.63 &0.25\\
 &    N-Bayes &0.41 &0.47 &0.14&0.44 &0.30 &0.09\\
     \hline
     \end{tabular}
    \caption{Classification accuracy for fluid intelligence based on dynamic connectivity using sliding window correlations.}
    \label{tab:dynamicFC}
\end{table}

\begin{table}[]
    \scriptsize
 \begin{tabular}{|l|lll|lll|lll|}
    \hline
   Without  &\multicolumn{3}{|c|}{CC200}&\multicolumn{3}{|c|}{AAL} &\multicolumn{3}{|c|}{HO} \\
   Age/ Sex  &MC&F1&I &MC&F1&I &MC&F1&I\\  
     \hline
     \hline
     DPL-SVM &\bf{0.14}&\bf{0.87}&\bf{0.72}&0.12&\bf{0.89}&{\bf 0.78}&\bf{0.15}&\bf{0.87}&\bf{0.74}\\
     Pen-SVM &0.16&0.84&0.70&0.13&0.87&0.76&0.17&0.86&0.73 \\
     Pen-Log &0.18&0.82&0.64&0.17&0.83&0.67&0.18&0.81&0.64\\
     Ind-SVM &0.30 &0.69 &0.39 &0.31 &0.69 &0.39&0.31&0.69&0.38\\
     N-Bayes &0.19 &0.81&0.64 &0.18 &0.82 &0.67 &0.24 &0.73&0.49\\
     \hline
     \hline
With &&& &&& &&& \\ 
Age/Sex  &MC&F1&I &MC&F1&I &MC&F1&I\\   
   \hline
       DPL-SVM & 0.14 & \bf{0.87} &\bf{0.73} &0.12&0.88&0.76&\bf{0.14}&0.87&\bf{0.74}\\
     Pen-SVM & 0.15&0.85&0.70&0.12&0.87&0.75&0.16&0.86&0.70 \\
     Pen-Log &0.17&0.83&0.67&0.17&0.83&0.69&0.17&0.82&0.67\\
      Ind-SVM &0.30 &0.70 &0.40 &0.31 &0.69 &0.39&0.31&0.69&0.38\\
     N-Bayes &0.17 &0.81 &0.67 &0.14 &0.83 &0.72 &0.23&0.75&0.56\\
     \hline
    \end{tabular}
    \caption{Classification accuracy for ADHD vs control using static FC (partial correlations) based on CNI data. }
    \label{tab:CNI}
\end{table}

{\noindent \underline{Results based on Multi-Session Analysis:}} The classification accuracy using static connectivity derived from the multi-session analysis involving both LR1 and LR2 scanning sessions, is consistently higher compared to the single-session analysis across all metrics and under all methods, as illustrated  in Table \ref{tab:static_results}. Moreover, the improvements under the multi-session analysis is almost always significantly better compared to the single-session analysis. These results clearly suggest that integrating  data across multiple scanning sessions leads to improvements in classification based on resting state connectivity, which is a novel finding of independent interest.

{\noindent \underline{Results based on Dynamic Connectivity:}} Table \ref{tab:dynamicFC} provides results for intelligence classification based on dynamic functional connectivity using a window size of 20. From amongst the three types of features extracted from the sliding window correlations, the manually extracted features defined in [19] yielded the best classification accuracy across all approaches. Hence, our analysis is able to validate the utility of the manually crafted metrics for dynamic functional connectivity proposed in [19] and is consistent in terms of reporting poor classification accuracy under ICA and PCA decomposition as also reported in [20]. Moreover, the ability of the resting state dynamic FC to distinguish individuals with different levels of fluid intelligence was greater in the LR1 scan compared to the LR2 scan, while the opposite was true for crystallized intelligence. These results are consistent with the results under the resting state static networks, and align with the definition of fluid and crystallized intelligence. 


As in the case for the static FC based classification, the proposed approach is shown to have higher accuracy compared to the competing methods for the majority of cases. However, the dynamic FC based classification results are consistently less accurate compared to resting state static FC results. This result illustrates that for rs-fMRI data, static FC is better suited for classifying intelligence levels compared to  dynamic connectivity. This suggests that incorporating temporal fluctuations in connectivity during the resting state scanning session may not provide advantages in classification compared to just using static connectivity in resting state experiments. This finding is complimentary to recent results in [20] that show good prediction accuracy for intelligence based on dynamic connectivity derived from task data in the HCP study.

\begin{figure}
    \centering
    \includegraphics[width=\linewidth]{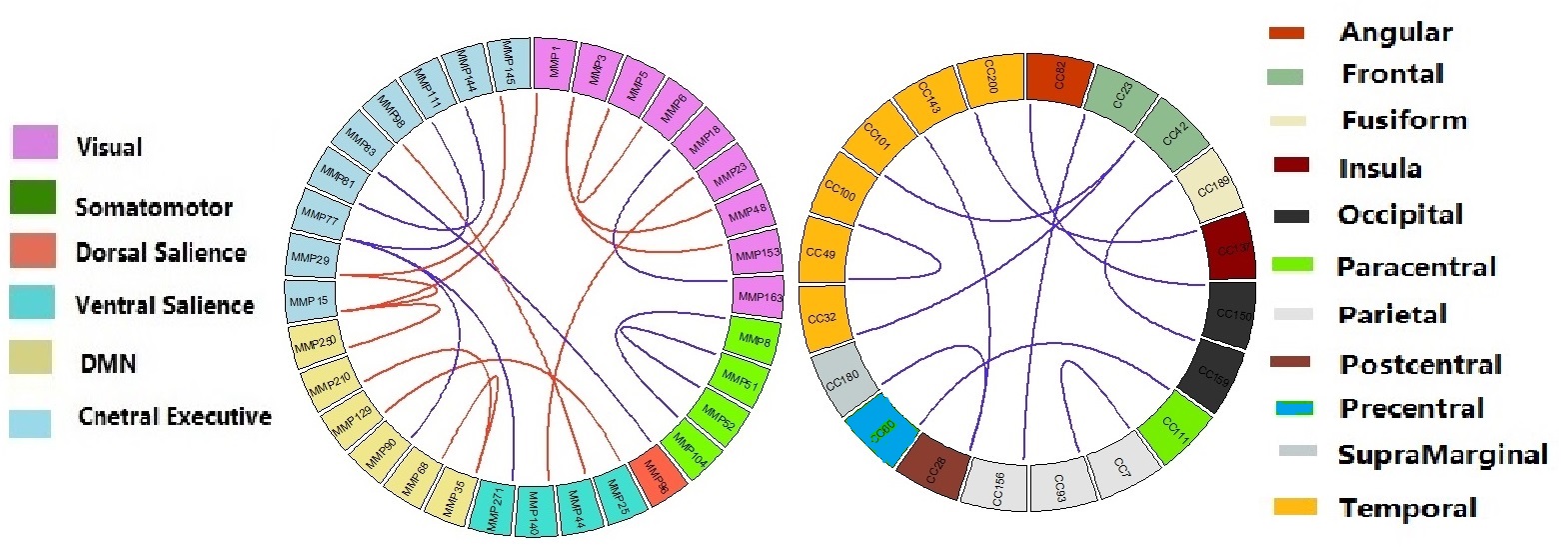}
    \caption{Circle plots denoting significant edges driving differences between groups, as inferred under the Bayesian SVM model. The left panel illustrates the significant edges corresponding to fluid (blue lines ) and crystallized (brown lines) intelligence in the HCP analysis. The right panel illustrates significant edges corresponding to ADHD versus NC classification for the CNI dataset based on the CC200 atlas.}
    \label{fig:cirplot}
\end{figure}

{\noindent \underline{Static Network Feature Selection Results:}} In terms of the feature selection, Figure \ref{fig:cirplot} illustrates edges that repeatedly show up as significant across multiple training/test splits under the LR1 analysis. For the crystallized intelligence classification, the significant edges were concentrated in the visual (3 W-M/2 B-M), central executive (2 W-M/3 B-M), default mode network (2 B-M/2 W-M), and ventral salience (3 B-M), where W-M and B-M refers to within module and between module connections respectively, indicating whether these significant edges were located entirely within the respective functional modules/networks, or whether they traversed multiple modules. The ROI corresponding to the parietal
occipital sulcus area in the central executive network with coordinates [99,53,109] had the highest number (3) of significant edges corresponding to crystallized intelligence, which is supported by previous evidence in literature [33]. On the other hand, the somatomotor (2 B-M) and central executive (2 B-M/3 W-M) modules has the greatest number of significant edges driving fluid intelligence classification. The ROI corresponding to the orbital and prefrontal cortex in the central executive network with coordinates [131,174,59] had the highest number (3) of significant edges corresponding to fluid intelligence classification that is supported by previous evidence in literature [34]. Our results solidify the important role of the central executive network in modulating intelligence.

\subsection{Analysis Results for CNI Data}

{\noindent \underline{Results based on Static FC:}} We report the classification results under static FC for the CNI data analysis in Table \ref{tab:CNI}, corresponding to the optimal network densities (18\% for CC200, 14\% for AAL, 20\% for HO atlases). We see improved classification performance under the proposed approach in the analysis excluding age and gender, across all the three different brain parcellation schemes, with the largest gains coming from the Craddock200 atlas having the largest number of brain regions. Moreover, all the three classification metrics are significantly higher under the proposed approach under the CC200 and HO atlas. In the analysis including age and gender, the proposed method still has a significantly higher classification accuracy in terms of the F1 and informedness score under the CC200 atlas, and significantly higher mis-classification accuracy and informedness score under the HO atlas. Out of the three parcellation schemes, the AAL atlas provides slightly better classification accuracy, which illustrates that the parcellation scheme can have an effect on the manner in which the network features are associated with the clinical outcome. We note that the classification accuracy reported in our analysis are comparable, and often higher compared to the results reported for the same dataset in [26]. In addition, the classification performance using dynamic connectivity based on sliding windows appear less accurate compared to the static connectivity results, that is consistent with our findings from the HCP study (not reported due to space constraints).

{\noindent \underline{Static Network Feature Selection Results:}} The significant edges are illustrated in Figure \ref{fig:cirplot}. Most of these edges appear within modules, i.e. traversing between different functional networks. The modules with the largest number of significant edges are seen to be concentrated in the temporal (4 W-M/ 1 B-M) and frontal regions (3 W-M). The ROIs with the largest number of significant edges were seen to lie in the Middle frontal gyrus in the orbital part, and the Inferior frontal gyrus in the 
opercular part, both in the left hemisphere.





\section{Conclusion}
Through this work, we provided concrete evidence that Bayesian SVMs with non-parametric priors on the coefficients are able to provide an unsupervised and automated approach for network-based classification. The proposed method results in statistically significant gains in the classification of intelligence levels and mental disorders compared to existing penalized classification  methods routinely used in literature, as well as parametric Bayesian approaches. In addition to being computationally efficient, the Bayesian approach is also able to perform inference for feature selection in order to identify important network edges contributing to differences in intelligence levels, in a reproducible manner. Our analysis revealed that the highest classification accuracy comes from using network edges derived from sparse precision matrices corresponding to static connectivity, and using data from multiple fMRI sessions in contrast to just one session that is standard in literature. On the other hand, classification using   dynamic connectivity as well as pairwise correlations for static connectivity (results not presented here) resulted in inferior classification accuracy compared to the analysis using static connectivity using sparse precision matrices. In our experience, we expect the classification accuracy to depend on several factors including the brain parcellation scheme adopted for the network, the heterogeneity in the population, as well as the nature of the binary phenotype. These may be explored in greater detail in future work.








\section*{REFERENCES}

\begin{enumerate}
\item Lukemire, J., Kundu, S., Pagnoni, G., \& Guo, Y. (2021). Bayesian joint modeling of multiple brain functional networks. Journal of the American Statistical Association, 116(534), 518-530.
\item Kundu, S., Ming, J., Nocera, J., \& McGregor, K. M. (2021). Integrative learning for population of dynamic networks with covariates. NeuroImage, 236, 118181.
\item Smith, S. M., Miller, K. L., Salimi-Khorshidi, G., Webster, M., Beckmann, C. F., Nichols, T. E., ... \& Woolrich, M. W. (2011). Network modelling methods for FMRI. Neuroimage, 54(2), 875-891.
\item Guha, S., \& Rodriguez, A. (2021). Bayesian regression with undirected network predictors with an application to brain connectome data. Journal of the American Statistical Association, 116(534), 581-593.
\item Ma, X., Kundu, S., \& Stevens, J. (2019). Semi-parametric Bayes Regression with Network Valued Covariates. arXiv preprint arXiv:1910.03772.
\item Meng, L., \& Xiang, J. (2018). Brain network analysis and classification based on convolutional neural network. Frontiers in computational neuroscience, 95.
\item Fan, L., Su, J., Qin, J., Hu, D., \& Shen, H. (2020). A deep network model on dynamic functional connectivity with applications to gender classification and intelligence prediction. Frontiers in neuroscience, 881.
\item Relion, J. D. A., Kessler, D., Levina, E., \& Taylor, S. F. (2019). Network classification with applications to brain connectomics. The annals of applied statistics, 13(3), 1648.
\item Supekar, K., Menon, V., Rubin, D., Musen, M., \& Greicius, M. D. (2008). Network analysis of intrinsic functional brain connectivity in Alzheimer's disease. PLoS computational biology, 4(6), e1000100.
\item Higgins, I. A., Guo, Y., Kundu, S., Choi, K. S., \& Mayberg, H. (2018). A differential degree test for comparing brain networks. arXiv preprint arXiv:1809.11098.
\item Du, Y., Fu, Z., \& Calhoun, V. D. (2018). Classification and prediction of brain disorders using functional connectivity: promising but challenging. Frontiers in neuroscience, 12, 525.
\item Chang, C., Kundu, S., \& Long, Q. (2018). Scalable Bayesian variable selection for structured high‐dimensional data. Biometrics, 74(4), 1372-1382.
\item Xiao, L., Stephen, J. M., Wilson, T. W., Calhoun, V. D., \& Wang, Y. P. (2019). A manifold regularized multi-task learning model for IQ prediction from two fMRI paradigms. IEEE Transactions on Biomedical Engineering, 67(3), 796-806.
\item Ma, X., \& Kundu, S. (2021). Multi-task Learning with High-Dimensional Noisy Images. arXiv preprint arXiv:2103.03370.
\item Bullmore, E., \& Sporns, O. (2009). Complex brain networks: graph theoretical analysis of structural and functional systems. Nature reviews neuroscience, 10(3), 186-198.
\item Chen, P. Y., Chen, C. L., Hsu, Y. C., \& Tseng, W. Y. I. (2020). Fluid intelligence is associated with cortical volume and white matter tract integrity within multiple-demand system across adult lifespan. NeuroImage, 212, 116576.
\item Yuan, P., Voelkle, M. C., \& Raz, N. (2018). Fluid intelligence and gross structural properties of the cerebral cortex in middle-aged and older adults: A multi-occasion longitudinal study. Neuroimage, 172, 21-30.
\item Shen, X., Finn, E. S., Scheinost, D., Rosenberg, M. D., Chun, M. M., Papademetris, X., \& Constable, R. T. (2017). Using connectome-based predictive modeling to predict individual behavior from brain connectivity. nature protocols, 12(3), 506-518.
\item Liu, J., Liao, X., Xia, M., \& He, Y. (2018). Chronnectome fingerprinting: Identifying individuals and predicting higher cognitive functions using dynamic brain connectivity patterns. Human brain mapping, 39(2), 902-915.
\item Sen, B., \& Parhi, K. K. (2020). Predicting biological gender and intelligence from fMRI via dynamic functional connectivity. IEEE Transactions on Biomedical Engineering, 68(3), 815-825.
\item Finn, E. S., Shen, X., Scheinost, D., Rosenberg, M. D., Huang, J., Chun, M. M., ... \& Constable, R. T. (2015). Functional connectome fingerprinting: identifying individuals using patterns of brain connectivity. Nature neuroscience, 18(11), 1664-1671.
\item Dadi, K., Rahim, M., Abraham, A., Chyzhyk, D., Milham, M., Thirion, B., ... \& Alzheimer's Disease Neuroimaging Initiative. (2019). Benchmarking functional connectome-based predictive models for resting-state fMRI. NeuroImage, 192, 115-134.
\item Chen, Y., Tang, Y., Wang, C., Liu, X., Zhao, L., \& Wang, Z. (2020). ADHD classification by dual subspace learning using resting-state functional connectivity. Artificial intelligence in medicine, 103, 101786.
\item Glasser, M. F., Coalson, T. S., Robinson, E. C., Hacker, C. D., Harwell, J., Yacoub, E., ... \& Van Essen, D. C. (2016). A multi-modal parcellation of human cerebral cortex. Nature, 536(7615), 171-178.
\item Smith, S. M., Beckmann, C. F., Andersson, J., Auerbach, E. J., Bijsterbosch, J., Douaud, G., ... \& WU-Minn HCP Consortium. (2013). Resting-state fMRI in the human connectome project. Neuroimage, 80, 144-168.
\item Schirmer, M. D., Venkataraman, A., Rekik, I., Kim, M., Mostofsky, S. H., Nebel, M. B., ... \& Chung, A. W. (2021). Neuropsychiatric disease classification using functional connectomics-results of the connectomics in neuroimaging transfer learning challenge. Medical image analysis, 70, 101972.
\item Friedman, J., Hastie, T., \& Tibshirani, R. (2008). Sparse inverse covariance estimation with the graphical lasso. Biostatistics, 9(3), 432-441.
\item Polson, N. G., \& Scott, S. L. (2011). Data augmentation for support vector machines. Bayesian Analysis, 6(1), 1-23.
\item Park, T., \& Casella, G. (2008). The bayesian lasso. Journal of the American Statistical Association, 103(482), 681-686.
\item Antoniak, C. E. (1974). Mixtures of Dirichlet processes with applications to Bayesian nonparametric problems. The annals of statistics, 1152-1174.
\item Walker, S. G. (2007). Sampling the Dirichlet mixture model with slices. Communications in Statistics—Simulation and Computation®, 36(1), 45-54.
\item Hutchison, R. M., Womelsdorf, T., Allen, E. A., Bandettini, P. A., Calhoun, V. D., Corbetta, M., ... \& Chang, C. (2013). Dynamic functional connectivity: promise, issues, and interpretations. Neuroimage, 80, 360-378.
\item Jung, R. E., \& Haier, R. J. (2007). The Parieto-Frontal Integration Theory (P-FIT) of intelligence: converging neuroimaging evidence. Behavioral and brain sciences, 30(2), 135-154.
\item Raz, N., Lindenberger, U., Ghisletta, P., Rodrigue, K. M., Kennedy, K. M., \& Acker, J. D. (2008). Neuroanatomical correlates of fluid intelligence in healthy adults and persons with vascular risk factors. Cerebral cortex, 18(3), 718-726.
\end{enumerate}

\end{document}